\def\noi{\noindent}
\def\be{\begin{equation}}
\def\ee{\end{equation}}
\def\bc{\begin{center}}
\def\ec{\end{center}}
\def\bea{\begin{eqnarray}}
\def\eea{\end{eqnarray}}
\def\ba{\begin{array}}
\def\ea{\end{array}}
\def\lb{\left\{}
\def\rb{\right\}}
\def\nn{\nonumber}
\def\um{\frac{1}{2}}

\documentclass[11pt]{article}

\usepackage[latin1]{inputenc}
\usepackage{amssymb}
\usepackage{amsfonts} 
\usepackage{amsthm} 
\usepackage{amsmath} 
\usepackage{epsfig}
\usepackage{hyperref}

\begin{document}

\begin{center}
{\LARGE {\bf Non-Canonical Perturbation Theory of \\ Non-Linear Sigma
Models}}
\end{center}

\bigskip
\bigskip

\centerline{{\sc V. Aldaya$^{1}$, M.
Calixto$^{2,1}$\footnote{Corresponding author:  Manuel.Calixto@upct.es}
and F.F. López-Ruiz$^{1}$}}

\bigskip
\bc {\it $^1$ Instituto de Astrof\'\i sica de Andaluc\'\i a
(IAA-CSIC), Apartado Postal 3004, 18080 Granada, Spain} \\ {\it
$^2$ Departamento de Matemática Aplicada y Estad\'\i stica,
Universidad Politécnica de Cartagena, Paseo Alfonso XIII 56, 30203
Cartagena, Spain}

 \ec

\bigskip

\begin{center}

{\bf Abstract}
\end{center}

\small We explore the $O(N)$-invariant Non-Linear Sigma Model (NLSM) in a
different perturbative regime from the usual relativistic-free-field one, by using
non-canonical basic commutation relations adapted to the underlying $O(N)$
symmetry of the system, which also account for the non-trivial (non-flat)
geometry and topology of the target manifold.

\setlength{\baselineskip}{12pt}

\begin{list}{}{\setlength{\leftmargin}{3pc}\setlength{\rightmargin}{3pc}}
\item
\end{list}

\normalsize

\noi PACS:  02.20.Tw,
03.70.+k, 
11.10.Lm, 
11.15.Bt, 
12.15.-y 

\setlength{\baselineskip}{14pt}


\section{Introduction}

From an abstract (mathematical) point of view, a Non-Linear Sigma
Model (NLSM) consists of a set of coupled scalar fields
$\pi^a(x^\mu), a=1,\dots,N,$ in a $D$-dimensional (Minkowski)
spacetime $M$ with coordinates $x^\mu, \mu=0,1,2,\dots,D-1$, and
action integral (we use the Einstein summation convention)
\begin{equation} S_\sigma(\pi,\partial_\mu\pi)=\int_M {\cal L}(\pi,\partial_\mu\pi){\rm d}^Dx=
\frac{\lambda}{2}\int_M
g_{ab}(\pi)\partial^\mu\pi^a\partial_\mu\pi^b\,{\rm d}^Dx,
\label{nlsmaction}\end{equation}
where $\partial^\mu=\eta^{\mu\nu}\partial_\nu,
\partial_\nu=\partial/\partial x^\nu$, $\eta={\rm diag}(+,-,\dots,-)$ is the Minkowski
metric and $\lambda$ a coupling constant. The field theory
(\ref{nlsmaction}) is called the NLSM with metric $g_{ab}(\pi)$
(usually a positive-definite field-dependent matrix). The fields
$\pi^a$ themselves could also be considered as the coordinates of
an internal Riemannian (target) manifold $\Sigma$ with metric
$g_{ab}$. This model proved to be relevant in String Theory where
$g_{ab}$ is the Einstein metric and $M$ is a two-dimensional
manifold named ``worldsheet''. An interesting case for us is that
in which $\Sigma$ is a Lie group manifold $G$, namely
$G=O(N)$, or a quotient (coset) space $G/H$ by a closed subgroup
$H$, namely $H=O(N-1)$ (see \cite{needs} for $G=U(N)$ and its
cosets $G/H$: complex projective, Grassmann and flag manifolds).

Apart from String Theory, the NLSM is
related to a great number of physical systems (see e.g.
\cite{Ketov} for a review). It was originally introduced to
describe pion dynamics in the theory of strong nuclear
interactions. Also, some particular two-dimensional
$O(N)$-invariant NLSM are used in connection to antiferromagnetic
spin chains, quantum Hall effect and superfluid helium-3. At a
more fundamental level, NLSM describes the dynamics of Goldstone
bosons in spontaneously broken field theories like the Standard
Model of electro-weak interactions. Recently we have proposed in
\cite{MPLAgauge-sigma} a Higgs-less mechanism to provide mass to
the electro-weak gauge vector bosons $W_\pm$ and $Z$ through a
coupling to a $U(2)$-invariant NLSM \emph{à la} Stueckelberg.
Actually, according to the widely named ``Equivalence Theorem''
\cite{Logitano,Bardeen}, a very heavy Higgs particle can be
eliminated from the broken symmetry programme in favor of
non-linear $\sigma$-like Goldstone bosons, so that the actual computation
of Feynman diagrams involving the longitudinal polarizations of
the (massive) vector bosons in electroweak interactions can be
resolved in terms of the corresponding diagrams among those scalar
fields. Unfortunately, the use of a NLSM Lagrangian has led to an
apparent insoluble dichotomy unitarity-renormalizability
\cite{dicotomia1,dicotomia2,dicotomia3,dicotomia4,Hurth} (see also the review \cite{StueckelbergRev}
and references therein). In fact, it is well known that the NLSM,
in general, suffers from unavoidable renormalizability problems
under the canonical quantization programme (see, for instance,
\cite{Ketov}). Canonical perturbation theory proceeds from the
action (\ref{nlsmaction}) by expanding
$g_{ab}(\pi)=\delta_{ab}+O(\pi^2)$ and perturbing around massless
fields fulfilling $\partial_\mu\partial^\mu\pi^a=0$. However, this
perturbation scheme is subject to criticism. On the one hand,
massless solutions do not exhaust the whole solution manifold, as
other (soliton, instanton, skyrmion) solutions are known to exist \cite{solitons}.
On the other hand, the non-trivial (non-flat) geometry and
topology of the target manifold $\Sigma$ and its possible
symmetries are not being taken into account or properly exploited.
Regarding the last issue, references like \cite{Faddeev} tackled
the perturbation theory for NLSM in terms of left-$G$-invariant
quantities $L_\mu(x)=g^{-1}(x)\partial_\mu g(x), g\in G$, which do
not depend on the parametrization of $G$.

As in Reference \cite{Isham}, we think that the trouble that
canonical quantization faces in dealing with systems bearing
non-trivial topology can be traced back to the ``tangent space''
approximation imposed at the very beginning of the (canonical)
quantization program. Already in the simple case of ``free''
particles moving on spheres, a proper quantization requires the
replacement of canonical commutators with the Lie-algebra
commutators of the Euclidean group \cite{Isham,sigmita}. We shall
pursue this idea in this letter and construct a perturbation
theory adapted to non-canonical (namely, Euclidean) commutation
relations for the particular case of $G=O(N+1)$ invariant NLSM with
$\Sigma=S^N=O(N+1)/O(N)$ the $N$-dimensional sphere. The
discretization of the corresponding equations of motion provides a
mechanical picture of the $O(N)$-invariant NLSM as a
$(D-1)$-dimensional lattice model of coupled rotators connected by
springs (see later on Sec. \ref{rotsec}). Actually, this
equivalence has already been considered in, for instance,
\cite{CCM1,CCM2} who used the so-called ``coupled cluster
method'' to approach this problem. Our aim here is to explore the
NLSM in a different regime from the usual (relativistic-free-field) one, by
using non-canonical basic commutation relations adapted to the
underlying $O(N)$ symmetry of the system.

\section{$O(N+1)$-Invariant NLSM \label{rotsec}}

The $O(N+1)$-invariant NLSM Lagrangian in (\ref{nlsmaction}) can
be obtained from the quadratic one
\begin{equation}
 {\cal L}(\vec\phi,\partial_\mu\vec\phi)=\frac{1}{2}\partial_\mu\vec\phi\cdot\partial^\mu\vec\phi, \,\,\,
\vec{\phi}=(\phi^1,\dots,\phi^{N+1})\in\mathbb
R^{N+1},\label{lagrangian}
\end{equation}
with the constraint $\vec\phi^2=\rho^2=$constant. A NLSM action of type (\ref{nlsmaction})
can be recovered from this Lagrangian by
eliminating $\phi^{N+1}$ in terms of
$\vec\pi=(\phi^1,\dots,\phi^N)$ or its stereographic
projection on $\mathbb R^N$. Here we shall work with $\vec\phi$ and keep in mind the constraint
$\vec\phi^2=\rho^2$. Using Lagrange
multipliers, the Euler-Lagrange equations of motion can be cast in the form:
\begin{equation}
 \square\vec\phi=\frac{\square\vec\phi\cdot \vec\phi}{\vec\phi^2}\vec\phi, \;\;\; \vec\phi^2=\rho^2, \label{eqmot}
\end{equation}
where $\square=\partial_\mu\partial^\mu$ denotes the d'Alembertian
or wave operator. For $N=3,D=2$, extra Wess-Zumino-Novikov-Witten
terms can be added to the Lagrangian (\ref{lagrangian}) so that
the model is known to be integrable since one is able to find an
infinite number of conserved quantities closing a Kac-Moody Lie
algebra (see e.g.\cite{Ketov}).


Let us briefly remind how the NLSM above also arises from a
$\phi^4$-theory by ``freezing out'' the Higgs field degree of
freedom (as in the above-mentioned Equivalence Theorem).
Actually, the term ``sigma'' makes reference to the original model
for an effective theory of the meson part of the low-energy
nuclear theory. The Lagrangian (\ref{lagrangian}) is modified by a
Higgs potential
\begin{equation}
 {\cal L}_g=\frac{1}{2}\partial_\mu\vec\phi\cdot\partial^\mu\vec\phi+\frac{g}{4}(\vec\phi^2-\rho^2)^2,\label{lagrangian2}
\end{equation}
with $g$ a positive constant. It is customary to write
\be \phi^{N+1}=\rho+\sigma, \; \phi^a=\pi^a, \,a=1,\dots,N,\label{spontaneous}\ee
for small perturbations  around
$\vec\phi_{(0)}=(0,\dots,\rho)$. The Lagrangian
(\ref{lagrangian2}) acquires then the following form in terms of $(\vec\pi,\sigma)$:
\begin{equation}
 {\cal L}_g=\frac{1}{2}\partial_\mu\vec\pi\cdot\partial^\mu\vec\pi
+\frac{1}{2}\partial_\mu\sigma\partial^\mu\sigma+\frac{m_\sigma^2c^2}{2}\sigma^2+\dots,\label{lagrangian3}
\end{equation}
which states that the $\sigma$-meson (Higgs field) has mass
$m_\sigma=\sqrt{2g}\rho/c$ whereas the $\pi$-mesons (pions) remain
massless. In fact, in the quantum theory, $\pi^a$ describe
Goldstone bosons associated with the spontaneous breakdown from
the $O(N+1)$ to the $O(N)$ symmetry for the choice of vacuum
$\langle 0|\phi^j|0\rangle=\rho\delta_{j,N+1}$.

The original NLSM Lagrangian  (\ref{lagrangian}) can be obtained from (\ref{lagrangian2})
by taking the limit $g\to\infty$ and
imposing $\vec\phi^2=\rho^2$ in order to keep the Lagrangian
finite except for an irrelevant c-number term. This corresponds to
$m_\sigma\to\infty$ so that the Higgs field degree of freedom has
been frozen (something physically reasonable since it has not been
experimentally observed  yet). Note that, even for large $g$, we
could always keep $m_\sigma$ finite by taking the vacuum
expectation value $\rho$ small. Actually, we are interested in
this regime in this article.

However, one should be very cautious in taking this limit, since
we are dramatically changing the topology of the field
configuration space. One can not guarantee in principle that the
procedure of \textit{perturbing} commutes with that of
\textit{constraining}. In this article we pursue the alternative
strategy of ``constraining and then perturbing'', instead of the
previous scheme of ``perturbing and then constraining''. Nowadays
it is widely known that \textit{constraining} does not actually
commute (in general) with \textit{quantizing} (see e.g.
\cite{quantconst1,quantconst2,quantconst3,quantconst4} for
discussions on non-equivalent quantizations of systems with
non-trivial configuration spaces).


Let us restrict ourselves, for the sake of simplicity, to the
$N=2$ case. The equations (\ref{eqmot}) can also be obtained as
Hamiltonian equations of motion
\be \dot{\vec{\phi}}=\frac{\partial \vec\phi}{\partial
t}\equiv\{\vec\phi,H\},\;\ddot{\vec{\phi}}=\frac{\partial^2
\vec\phi}{\partial
t^2}\equiv\{\dot{\vec\phi},H\}=\{\{\vec\phi,H\},H\},\label{eqmot2}\ee
for the Hamiltonian
\be H=\um\int d^{D-1}x\left(\frac{\vec
L^2(x)}{\rho^2}+c^2(\vec\nabla\vec\phi(x))^2\right),\label{hamiltonian}\ee
and the basic equal-time Euclidean (non-canonical) Poisson
brackets
\be \lb L^i(x),L^j(y)\rb ={\epsilon^{ij}}_k
L^k(x)\delta(x-y),\;\;\; \lb L^i(x),\phi^j(y)\rb
={\epsilon^{ij}}_k \phi^k(x)\delta(x-y),\label{euclideancom}\ee
where $\vec L\equiv \vec\phi\wedge\dot{\vec\phi}$,
$(\vec\nabla\vec\phi)^2\equiv
\partial_j\phi_k\partial^j\phi^k$, ${\epsilon^{ij}}_k$ is the
antisymmetric symbol and we have introduced the wave velocity $c$
when setting $x^0=ct$ for later convenience. Actually, if
(\ref{euclideancom}) are taken as abstract Poisson brackets, with
$\vec L$ not necessarily related to $\vec\phi$, then the equations
(\ref{eqmot2}) generalize (\ref{eqmot}) by introducing an extra
term
\begin{equation}
 \square\vec\phi=\frac{\square\vec\phi\cdot \vec\phi}{\vec\phi^2}\vec\phi+
 \frac{\vec L\cdot\vec\phi}{\vec\phi^4}\vec L
  \label{eqmot3}
\end{equation}
which could not vanish when $\vec L\cdot\vec\phi\not=0$, a
situation which arises when ``magnetic monopoles'' are present and
$\vec L$ is not necessarily perpendicular to $\vec\phi$. We shall
restrict ourselves to the case $C_1=\vec L\cdot\vec\phi=0$, which
is compatible with the Poisson brackets (\ref{euclideancom}) and
the constraint $C_2=\vec\phi^2=\rho^2$, since both $C_1$ and $C_2$
are the natural Casimir operators for the Euclidean group.

Perturbing around $\vec\phi_{(0)}=(0,0,\rho)$ as in
(\ref{spontaneous}), for fixed $\vec\phi^2=\rho^2$, can be
interpreted as a ``group contraction'', which drastically changes
the topology of the system. Indeed, this perturbation theory has
sense for $\rho\gg 1$. Making the change (\ref{spontaneous}) in
the last Poisson bracket of (\ref{euclideancom}) and taking the
limit $\rho\to\infty$, keeping $\varphi^{1,2}\equiv
\pi^{1,2}/\rho$ finite, we recover the canonical Poisson brackets:
\be  \lb L^i(x),\varphi^j(y)\rb ={\epsilon^{ij}}_3 \delta(x-y),
\;\; i,j=1,2\label{euclideancom2}\ee
which state that $(\varphi^1,\varphi^2)$ and $(L^2,-L^1)$ are
couples of canonically-conjugated  variables. Therefore, standard
(canonical) perturbation theory has sense for large values of
$\rho$, which loses information about the (compact) topology of
the system. As already commented, we are interested in the other
regime $\rho\ll 1$.

\section{Classical non-canonical perturbation theory}

A solution of $\square\vec\phi=-m^2\vec\phi$, for any constant
$m$, is also a solution of (\ref{eqmot}). However, only for
massless fields, $m=0$, the constraint $\vec\phi^2=\rho^2$ is also
satisfied. At least at the quantum level, standard perturbation
theory proceeds by considering scattering of massless fields
$\vec\phi$ \cite{Ketov}. However, at the classical level, we know
that there are more solutions of (\ref{eqmot}) than massless
solutions. In fact, as showed long time ago in \cite{solitons},
the configuration space of a NLSM breaks up into an (infinite)
number of components. Indeed, finite energy requires boundary
conditions like (for instance) $\vec\phi(x)=(0,0,\rho)$ as
$\|x\|\to\infty$, which means a one-point compactification of
$\mathbb R^{D-1}$ by $S^{D-1}$. Thus, if two fields $\vec\phi$ and
$\vec\phi'$ belong to different homotopical classes
$\Pi_{D-1}(S^N)$, then they can not be continuously deformed
(evolved) one into the other. In particular, one can find
(solitonic) solutions that are not wave packets of massless
solutions.

Instead of perturbing around massless solutions, we shall adopt the following
splitting of the Hamiltonian (\ref{hamiltonian})
\be H=H_0+V,\;\; H_0=\um\int d^{D-1}x \frac{\vec L^2(x)}{\rho^2},\;\;
V=\frac{c^2}{2}\int d^{D-1}x (\vec\nabla\vec\phi(x))^2,\ee
and consider $V$ as a perturbation  for either small $c$ or
$\|\vec\phi\|=\rho \ll 1$ (with $c$ arbitrary).

\begin{figure}[htb]
\begin{center}
\includegraphics[height=4cm,width=9cm]{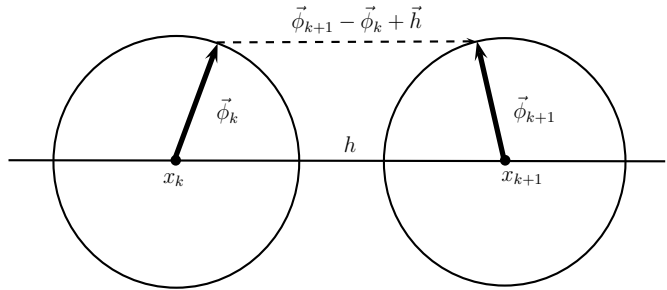}
\end{center}
\caption{Rotators in a lattice coupled by springs}\label{coupledrotors}
\end{figure}

Given an initial condition on a Cauchy hypersurface,
$\vec\phi(t_0,x)=\vec\phi_0(x)$ and
$\dot{\vec\phi}(t_0,x)=\dot{\vec\phi}_0(x)$, the general solution to
(\ref{eqmot}) can be formally written as:
\be \vec\phi(t,x)=e^{(t-t_0)\{\cdot,H\}}\vec\phi_0(x)=U(t-t_0)\vec\phi_0(x),\label{evol}\ee
where $\{\cdot,H\}$ stands for the Liouvillian operator and $U(t-t_0)=e^{(t-t_0)\{\cdot,H\}}$ for the evolution operator.
Actually, we can exactly integrate the ``free'' evolution as:
\bea
\vec\phi^{(0)}(t,x)&\equiv& e^{(t-t_0)\{\cdot,H_0\}}\vec\phi_0(x)=U_0(t-t_0)\vec\phi_0(x)\nn\\ &=&
\cos\left(\sqrt{\frac{\vec L^2(x)}{\rho^4}}\,(t-t_0)\right)\vec\phi_0(x)+
\frac{1}{\sqrt{\frac{\vec L^2(x)}{\rho^4}}}\sin\left(\sqrt{\frac{\vec L^2(x)}{\rho^4}}\,(t-t_0)\right)
\dot{\vec\phi}_0(x).\label{freesolution}\eea
We shall let the wave velocity $c$ to take arbitrary values, as we
want our perturbation theory to be valid for relativistic fields
too. We have already justified the interesting regime $\rho\ll 1$
(small vacuum expectation value) in which the Higgs mass
$m_\sigma$ would remain finite while $\vec\phi^2\simeq \rho^2$, so
that the Higgs field degree of freedom is almost frozen. In order
to gain more physical intuition on this limit, let us use the
following mechanical picture of coupled small rotators (see Figure
\ref{coupledrotors}).

Without loss of generality, we can restrict ourselves to $D=2$,
consider the lattice $x_k=kh, k\in\mathbb Z$, for some (small)
step $h$, and write $\vec\phi_k(t)=\vec\phi(t,x_k)$ for the vector
position of the rotator in the place $x_k$. Rotators are connected
by identical springs of constant $\kappa$ and zero natural length
so that the elastic potential energy between two consecutive
rotators is
\be
V_{k+1,k}=\um \kappa (\vec\phi_{k+1}-\vec\phi_k+h(1,0))^2.\ee
Taking the limit $h\to 0$, keeping $\kappa h\equiv c^2$  finite, we have that the total elastic potential energy is
\be \sum_{k=-\infty}^\infty V_{k+1,k}=\um c^2
\sum_{k=-\infty}^\infty
 h\frac{(\vec\phi_{k+1}-\vec\phi_k+h(1,0))^2}{h^2}\to \um
c^2\int_{-\infty}^\infty dx ((\partial_x\vec\phi)^2+2\partial_x
\phi^1).\ee
which gives the desired result up to a boundary term.

Contrary to the (unconstrained) Klein-Gordon field (as a model of
coupled oscillators), the elastic potential energy $V$ can be made
arbitrarily small for NLSM fields (as a model of coupled rotators)
by taking $\rho\ll 1$, even in rigid media ($c$ arbitrary). In
other words, unlike a NLSM, a Klein-Gordon field could never be
seen as an infinite set of weakly coupled oscillators unless
inside soft media ($c\ll 1$) where it takes a long time for the
wave to propagate. That is, here we have the vacuum expectation
value $\rho$ as an extra perturbation parameter to play with.

Although Dyson series are conventionally designed for quantum
perturbation theory, we shall briefly remind the subject here in a
classical setting. Dyson series takes advantage of the exact
solvability of $H_0$, with exact solution (\ref{freesolution}), to
provide a perturbation series in $V$. The evolution operator
(\ref{evol}) is decomposed as:
\be U(t,t_0)=U_0(t)\underbrace{U_0(-t)U(t-t_0)U_0(t_0)}_{
U_I(t,t_0)}U_0(-t_0),\ee
where $U_I(t,t_0)$ is the evolution operator in the interaction
image. Let us set $t_0=0$ for simplicity. After a little bit of
algebra, one can see that
\be \frac{\partial}{\partial t}U_I(t)=\hat V(t)U_I(t), \;\; \hat
V(t)\equiv U_0(t)\{\cdot, V(\vec\phi)\}U_0(-t)=\{\cdot,
V(\vec\phi^{(0)}(-t))\},\label{Vint}\ee
with $\vec\phi^{(0)}(t)$ given by $(\ref{freesolution})$ (note the
time inversion). This formula can be recursively integrated as:
\be U_I(t)=I+\int_0^t d\tau \hat V(\tau)+\int_0^t d\tau
\int_0^\tau d\tau' \hat V(\tau)\hat V(\tau')+\dots\label{evolint}\ee
In order to test the perturbation procedure, let us consider the
exactly solvable case $N=1, D=2$. On the one hand, if we
parametrize the field $\vec\phi=(\phi^1,\phi^2)$ in polar
coordinates $\phi=\rho e^{i\theta}$, then (\ref{eqmot}) reduces to
a massless Klein-Gordon equation for $\square\theta=0$. On the
other hand, we can compute order by order:
\be
\phi(t,x)=U(t)\phi_0(x)=U_0(t)U_I(t)\phi_0(x)=U_0(t)\phi^{(I)}(t,x),\label{solutot}\ee
with
\be \phi^{(I)}(t,x)=U_I(t)\phi_0(x)=\phi_0(x)+\int_0^t d\tau
\{\phi_0(x), V(\phi^{(0)}(-\tau))\}+\dots, \ee
where $V(\phi)=\frac{c^2}{2}\int_{-\infty}^\infty
dx\partial_x\phi\partial_x\bar\phi$ and Poisson brackets are
computed at $\tau=0$. Taking into account that
\be \phi^{(0)}(\tau,x)=U_0(\tau)\phi_0(x)=e^{i\tau L(x)/\rho^2}\phi_0(x)\label{freevol}\ee
with $L=\phi^1\dot\phi^2-\phi^2\dot\phi^1={\rm Im}(\bar\phi\dot\phi)$, and that
\be \{L(x),\phi(y)\}={-i}\phi(x)\delta(x-y),
\,\{\Psi(L(x)),\phi(y)\}=\Psi'(L(x))\{L(x),\phi(y)\},\label{afincom}\ee
for any derivable function $\Psi$ of $L$, we can compute at first
order:
\bea \phi_0^{(1)}(\tau,x)&\equiv&\{\phi_0(x),
V(\phi^{(0)}(-\tau))\}\nn\\
&=&\frac{c^2}{2}\left(\partial^2_{xx}\phi^{(0)}(-\tau,x)\tau
e^{i\tau L(x)/\rho^2}
-\partial^2_{xx}\bar\phi^{(0)}(-\tau,x)\frac{\tau}{\rho^2}
e^{-i\tau L(x)/\rho^2}\phi_0^2(x)\right)\nn\\ &=&
\frac{ic^2\tau}{\rho^2} (\rho^2\partial^2_{xx}\theta_0(x)-\tau
\partial^2_{xx}L(x))\phi_0(x),\eea
where we have put $\phi_0(x)=\rho e^{i\theta_0(x)}$. Therefore,
\bea \phi^{(I)}(t,x)&=&U_I(t)\phi_0(x)=\phi_0(x)+\int_0^t d\tau
\phi_0^{(1)}(\tau,x) +\dots\\ &=& \phi_0(x)\left(1+ \frac{ic^2
t^2}{2\rho^2} (\rho^2\partial^2_{xx}\theta_0(x)-\frac{2}{3}t
\partial^2_{xx}L(x))+\dots\right).\eea
The last step in (\ref{solutot}), i.e.
$\phi(t)=U_0(t)\phi^{(I)}(t)$, is easily performed by replacing
$\phi_0(x)$ by $\phi^{(0)}(t,x)$  (and $\theta_0(x)$ by
$\theta^{(0)}(t,x)=\theta_0(x)+tL(x)/\rho^2$) everywhere in
$\phi^{(I)}(t,x)$. That is:
\be \phi(t,x)=U_0(t)\phi^{(I)}(t,x)=\phi^{(0)}(t,x)\left(1+
\frac{ic^2 t^2}{2} (\partial^2_{xx}\theta_0(x)+\frac{1}{3\rho^2}t
\partial^2_{xx}L(x))+\dots\right).\ee
One can check that, at this
order, the perturbative solution coincides with the exact solution
$\phi(t,x)=\rho e^{i\theta(t,x)}$ where
\be \theta(t,x)=\cos(c t\partial_x)\theta_0(x)+\frac{\sin(c
t\partial_x)}{c\partial_x}\dot\theta_0(x).\ee
Inside the discrete, mechanical picture depicted in Figure
\ref{coupledrotors}, the appearance of second order spatial
derivatives $\partial^2_{xx}$ at first order in perturbation
theory means that the interaction propagates from one point $x_k$
to its nearest neighbors $x_{k+1}$ and $x_{k-1}$ at this order. In
order to account for a longer range propagation we should go to
higher orders in perturbation theory.

\section{Quantum non-canonical perturbation theory}

In quantum field theory the fields $\phi(x)$ and $L(x)$ are
promoted to the quantum operators $\hat\phi$ and $\hat L$,
respectively, and the Poisson brackets (\ref{euclideancom}) and
(\ref{afincom}) are promoted to the (non-canonical) commutators
(we shall keep restricting ourselves to $N=1, D=2$, for
simplicity):
\be \left[\hat
L(x),\hat\phi(y)\right]=\hbar\hat\phi(x)\delta(x-y),\;\;
\left[\hat
L(x),\hat{\phi}^\dag(y)\right]=-\hbar\hat{\phi}^\dag(x)\delta(x-y),\ee
where we have introduced $\hbar$ just to account for quantum
corrections and $\rho^2=\hat{\phi}(x)\hat{\phi}^\dag(x)$ gets the
necessary dimensions to render the Hamiltonian with energy
dimensions. Let us consider the lattice picture of our field model
and write $\hat\phi(x_k)=\hat\phi_k$ and $\hat L(x_k)=\hat L_k$.
The Hilbert space ${\cal H}_k={\rm Span}(|n_k\rangle,
n_k\in\mathbb Z)$ of a single rotator at position $x_k$ is spanned
by the (normalized) eigenstates $|n_k\rangle$ of the angular
momentum $\hat L_k$, that is:
\be
\hat L_k|n_k\rangle=\hbar n_k|n_k\rangle.\ee
The operators $\hat{\phi}_k$ and $\hat{\phi}^\dag_k$ act on $|n_k\rangle$ as ladder operators, namely:
\be
\hat{\phi}_k|n_k\rangle=\rho|n_k+1\rangle,\;\; \hat{\phi}^\dag_k|n_k\rangle=\rho|n_k-1\rangle.\ee
The total Hilbert space ${\cal H}$ of our lattice quantum field
theory will be the direct product ${\cal
H}=\bigotimes_{k\in\mathbb Z}{\cal H}_k$. The total Hamiltonian
operator is
\be
\hat H=\hat H_0+\hat V, \;\; \hat H_0=\frac{\omega}{2\hbar}\sum_{k=-\infty}^\infty \hat L_k^2,\; \;
\hat V(\hat\phi)=-\kappa\sum_{k=-\infty}^\infty {\rm Re}(\hat\phi_{k+1}\hat{\phi}^\dag_k),\ee
where we have discarded a c-number addend in $\hat V$ and we have introduced a frequency $\omega\equiv h\hbar/\rho^2$.
In order to write the evolution operator in the interaction image $U_I(t)$, we need to
evolve $\phi_k$ with the free evolution operator $U_0(t)=e^{-\frac{it}{\hbar}\hat H_0}$:
\be
\hat\phi_k^{(0)}(t)=U_0(-t)\hat\phi_kU_0(t)=\sum_{m=0}^\infty \frac{({-it}/{\hbar})^m}{m!}[\hat\phi_k,\hat H_0]^{(m)},\ee
where we denote the multiple commutator:
\be
[\hat\phi_k,\hat H_0]^{(m)}\equiv [[\hat\phi_k,,\hat H_0],\stackrel{m}{\dots},\hat H_0].\label{multiplecom}\ee
The quantum commutator introduces new ordering problems with
respect to the classical Poisson bracket. For standard creation
$\hat a^\dag_k$ and annihilation $\hat a_k$ operators,  Wick's
theorem provides a useful tool for writing arbitrary products of
$\hat a^\dag_k$ and $\hat a_l$ in terms of normal ordered
products. Here we have to deduce a new Wick-like theorem in order
to write arbitrary products of the non-canonical operators $\hat
L_k$ and $\hat\phi_l$. If we choose by convention to write all
$\hat L$'s to the left of all $\hat\phi$'s, then the multiple
commutator  (\ref{multiplecom}) acquires the following form:
\be
[\hat\phi_k,\hat H_0]^{(m)}=\frac{(-1)^m \omega^m}{2^m}\left(\sum_{l=0}^m c_{m,l}
\hbar^{l} \hat L^{m-l}_k\right)\hat \phi_k={(-1)^m \omega^m}(\hat L_k^m+q.c.)\hat \phi_k,\ee
where $q.c.$ stands for ``quantum corrections''. The Wick-like
numerical coefficients  $c_{m,l}$ are given by $c_{m,0}=2^m, \,
c_{m,m}=(-1)^m$ and the recurrence
$c_{m,l}=2c_{m-1,l}-c_{m-1,l-1}$. Therefore \be
\hat\phi_k^{(0)}(t)=U_0(-t)\hat\phi_kU_0(t)=(e^{it\omega\hat
L_k/\hbar}+q.c.)\hat\phi_k,\ee
coincides with the classical expression (\ref{freevol}) except for
quantum corrections.  Actually, we shall be able to sum up all
quantum corrections in some particular cases (see later) by
noticing that
\be\sum_{l=0}^m c_{m,l}=1,\;\; \forall m=0,1,2,\dots\label{sumproperty}\ee

The evolution operator in the interaction image (\ref{evolint}) is
given in terms of
\be
\hat V(\hat\phi^{(0)}(-\tau))=-\kappa\sum_{q=-\infty}^\infty
{\rm Re}\left((e^{-i\tau\omega\hat L_{q+1}/\hbar}+q.c.)\hat\phi_{q+1}\hat{\phi}^\dag_q(e^{i\tau\omega\hat
L_q/\hbar}+q.c.)\right).\label{VIop}\ee
In order to describe the new perturbation scheme, let us consider an initial state (at time $t=0$)
\be
|\{n\}\rangle=\otimes_{q\in\mathbb Z} |n_q\rangle.\ee
The probability amplitude of observing $|\{n'\}\rangle$ as a final state after time $t$ is
given by the $S$ matrix element:
\be
S_{n,n'}(t)=\langle \{n'\}|U(t)|\{n\}\rangle=\langle \{n'\}|U_0(t)U_I(t)|\{n\}\rangle=
e^{it \frac{\omega}{2}\sum_{k=-\infty}^\infty (n'_k)^2}\langle \{n'\}|U_I(t)|\{n\}\rangle.\ee
The total angular momentum $\hat L=\sum_k \hat L_k$ is conserved at all orders in perturbation theory since $[\hat L,\hat V]=0$. This means that
\be
\sum_{k=-\infty}^\infty n_k\not=\sum_{k=-\infty}^\infty n'_k\Rightarrow  S_{n,n'}(t)=0.\ee
The interaction potential (\ref{VIop}) is of short range, that is,
$\hat V$ is not able to carry one quantum of angular momentum from
position $k$ to $l$ until $|k-l|$-th order in perturbation theory.
More precisely, considering an initial state of the form \be
|\{\delta_k\}\rangle=\otimes_{q\in\mathbb Z}
|\delta_{k,q}\rangle,\ee
we can compute the probability amplitude of observing
$|\{\delta_l\}\rangle$  as a final state after time $t$ at all
orders:
\be S_{\delta_k,\delta_l}(t) =e^{it \omega/2}\sum_{n=0}^\infty
\left(\frac{it\kappa\rho^2}{2\hbar}\right)^n\frac{1}{n!}\sum_{s=0}^{n}\binom{n}{s}\delta_{l,k-n+2s},\ee
%
where we have made use of (\ref{sumproperty}) at some stage. Note
that  perturbation theory is dictated by both: $\kappa$  and/or
$\rho$.

Instead of the angular momentum eigenstates $|n_k\rangle$
we could also have used field eigenstates
\be |\zeta_k\rangle\equiv\sum_{n=-\infty}^\infty
\zeta_k^n|n\rangle,\;\; |\zeta_k|=1,\ee
for which $\hat\phi_k|\zeta_k\rangle=\rho\zeta_k|\zeta_k\rangle$,
$\hat\phi_k^\dag|\zeta_k\rangle=\rho\zeta_k^{-1}|\zeta_k\rangle$ and $\hat
L_k|\zeta_k\rangle=\hbar\zeta_k\partial_{\zeta_k}|\zeta_k\rangle$.
Moreover, going from $N=1$ to arbitrary $N$ can be accomplished by
replacing $|n_k\rangle$ with hyper-spherical harmonics. For $N=2$, the
usual spherical harmonics are given in terms of homogeneous polynomials of
degree $j$ in $\vec\phi$:
\be
Y^j_m(\vec\phi_k)=\sum_{\ba{l} a_q=1,2,3\\ q=1,\dots,j\ea}
\xi^{(m)}_{a_1,\dots,a_j}\phi^{a_1}_k\dots \phi^{a_j}_k, \ee
where $\xi^{(m)}_{a_1,\dots,a_j}$ are the complex components of a
symmetric and traceless tensor \cite{Louck}. The angular momentum
operator at place $x_k$ is then given by $\hat
L_k^a=\hbar{\epsilon^{ab}}_c \phi_k^c\partial_{\phi_k^b}$, as
usual.

\section{Conclusions}

The usual perturbation theory for relativistic fields is designed for
small deviations from the free (Klein-Gordon or Dirac) fields. The data analysis of detectors
in particle colliders is also intended for this purpose. However,
fields of NLSM-type can be found in a strongly-interacting regime
($\rho\ll 1$) which does not fit into this picture. This leads us to
reconsider the perturbation theory and renormalizability of the NLSM.

In this paper we have considered a non-canonical approach to the
perturbation theory of the $O(N)$-invariant NLSM which accounts for the
non-trivial (non-flat) geometry and topology of the target manifold
$\Sigma$ and takes advantage of the underlying symmetries of the system.
This scheme can also be adapted to other $G$-invariant NLSM.

\section*{Acknowledgements}
Work partially supported by the Fundación Séneca (08814/PI/08),
Spanish MICINN (FIS2008-06078-C03-01) and Junta de Andaluc\'\i a
(FQM219, FQM1951). F.F.
López-Ruiz thanks C.S.I.C. for an I3P grant. M. Calixto thanks the ``Universidad Politécnica de Cartagena''
and C.A.R.M.  for the award  ``Intensificación de la Actividad
Investigadora 2009-2010''.

\end{document}